\documentclass[12pt]{iopart}
\usepackage{iopams}  

\usepackage{graphicx}% Include figure files
\usepackage{dcolumn}% Align table columns on decimal point
\usepackage{bm}% bold math

\begin{document}

\newcommand{\cm}{cm$^{-1}$} \newcommand{\A}{\AA$^{-1}$} \newcommand{\Q}{\mathbf{Q}} 

\title[Nonlocal protons and deuterons]{Nonlocal protons and deuterons oppose to disorder: A single-crystal neutron diffraction study of  KH$_{0.76}$D$_{0.24}$CO$_3$ and a theoretical framework}

\author{Fran\c{c}ois Fillaux\dag\ and Alain Cousson\ddag}
\address{\dag\ LADIR-CNRS, UMR 7075 Universit\'{e} P. et M. Curie, 2 rue Henry Dunant, 94320 Thiais, France}
\address{\ddag\ Laboratoire L\'{e}on Brillouin, CEA-CNRS, C.E. Saclay, 91191 Gif-sur-Yvette cedex, France}
\ead{fillaux@glvt-cnrs.fr}

\date{\today}

\begin{abstract}
We show with neutron diffraction that a single-crystal of potassium hydrogen deuterium carbonate at room temperature, namely KH$_{0.76}$D$_{0.24}$CO$_3$, is isomorphous with the KHCO$_3$ and KDCO$_3$ derivatives. Protons and deuterons are not disordered particles located at definite sites and Bragg peaks are best fitted with separate H and D sublattices. We propose a theoretical framework for nonlocal observables and macroscopic states compatible with the crystal field. 
\end{abstract}


%\keywords{Neutron diffraction; Isotope mixture; disorder; Nonlocality; Quantum superposition; Hydrogen bonding}

\submitto{\JPCM}

\maketitle

\section{Introduction}

Defect-free crystals are macroscopic quantum systems with spatial periodicity. This is attested by electronic states and, for nuclei, by discrete phonon spectra observed at any temperature below melting or decomposition. However, the quantum nature of nuclei is overlooked in many cases. For example, in crystals containing O--H$\cdots$O hydrogen bonds, when the coexistence of two configurations for protons, say $\mathrm{O1-H}\cdots \mathrm{O2}$ and $\mathrm{O1}\cdots \mathrm{H-O2}$, is conceived of as ``disorder'' of particles with definite positions and momenta. This is at variance with vibrational spectra showing that the translational invariance of the lattice and the quantum nature of dynamics are not destroyed, even on the short time scale of proton dynamics ($\sim 10^{-12}-10^{-14}$ s). Such crystals should be therefore represented by macroscopic states for nonlocal observables. 

This conflict between disorder and macroscopic states parallels the dichotomy of interpretation of quantum mechanics applied to macroscopic systems \cite{Laloe,Leggett}. At the microscopic level, a quantum superposition indicates a lack of definiteness of outcome evidenced by interferences. Extrapolated to the macroscopic level of everyday life, this leads to Schr\"{o}dinger's Cat in a superposition of $|alive\rangle$ and $|dead\rangle$ states. However, we know that a cat is either dead or alive, so we are naturally inclined to think that a macroscopic object which has available to it two or more macroscopically distinct states is in one of these states for each measurement of the ensemble \cite{Leggett1,LG,Leggett5}. For hydrogen bond protons, this means that dynamics at the microscopic level of a double-well are quantum in nature (tunnelling), whereas at the macroscopic level of a crystal, protons which have available to them distinct sites are in one of the many possible configurations for each measurement. 

For open systems, this dichotomy of interpretation can be legitimated by decoherence stipulating that an initial superposition state should lose its ability to exhibit quantum interferences via interaction with the environment \cite{Zurek}. However, since the quantum theory does not predict any dividing line between quantal and classical regimes, macroscopic quantum behaviour is possible for systems decoupled from, or very weakly coupled to, the surroundings \cite{ACAL}. In principle, there is no upper limit in size, complexity, and temperature, beyond which such systems should be doomed to classicality. 

The crystal of potassium hydrogen carbonate (KHCO$_3$) is a counterexample to the dichotomy of interpretation of quantum mechanics at the macroscopic level. Neutron diffraction and vibrational spectroscopy provide positive evidences of macroscopic superposition states for protons featuring quantum correlations, at the scale of Avogadro's constant, and about room temperature \cite{Fil3,IF,FCKeen,FCG2,Fil7,FCG4}. The particle-like representation of protons possessing properties in their own right (position, momentum, spin) must be abandoned. Nonlocal observables and macroscopic wave functions are best appropriate to the translation invariant symmetry of the crystal. These states are decoherence-free and there is no transition to the classical regime, as long as the crystal structure is not disrupted. 

The question at issue in the present paper is as to whether or not macroscopic states are destroyed by static disorder in a mixed crystal, KH$_{p}$D$_{1-p}$CO$_3$, as grown from a water solution containing H$_2$O, HDO, D$_2$O molecules. In this crystal, H and D nuclei are not mobile and there is no segregation into separate domains. Our purpose is to determine with neutron diffraction the impact of static disorder to the crystal symmetry. We are not aware of such measurements ever published for this system. 

In Section \ref{sec:2}, we report neutron diffraction measurements showing that the crystal of KH$_{0.76}$D$_{0.24}$CO$_3$ is isomorphous with KHCO$_3$ and KDCO$_3$. This is at variance with static disorder. Data are best fitted with distinct H and D sublattices. We propose, in Section \ref{sec:3}, a theoretical framework for superposed proton and deuteron states. 

\section{\label{sec:2}Crystal structures}

\subsection{\label{sec:21}KHCO$_3$ and KDCO$_3$}

\begin{figure}[!hbtp]
\begin{center}
\includegraphics[scale=.25]{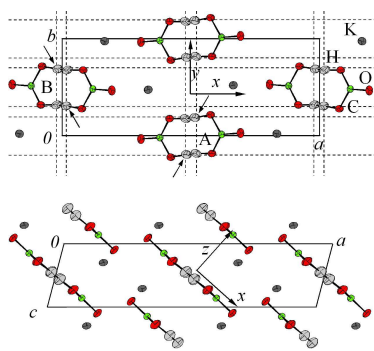}
\end{center}
\caption{\label{fig:1} Schematic view of the crystalline structure of KHCO$_3$ at 300 K. The arrows point to the sites occupied at low temperature (the $L$ configuration, see text). Dashed lines through protons are guides for the eyes. The ellipsoids represent 50\% of the probability density for nuclei.}
\end{figure}

Up to room temperature, these monoclinic crystals are isomorphous: space group $P2_1/a$ (C$_{2h}^5$) with four equivalent KHCO$_3$ or KDCO$_3$ entities per unit cell (Fig. \ref{fig:1}) \cite{TTO1,TTO2}. They are made up of centrosymmetric dimers (HCO$_3^-)_2$ or (DCO$_3^-)_2$ linked by moderately strong hydrogen (deuterium) bonds, with lengths $R_{\mathrm{OHO}} \approx  2.58$ \AA\ ($R_{\mathrm{ODO}} \approx  2.61$ \AA). Dimers lie practically in (103) planes and hydrogen (deuterium) bonds are virtually parallel to each other. These crystals are uniquely suited to probing dynamics along directions $x,\ y,\ z,$ parallel to the stretching ($\nu$OH/$\nu$OD), the in-plane bending ($\delta$OH/$\delta$OD), and the out-of-plane bending ($\gamma$OH/$\gamma$OD) vibrations, respectively. 

\begin{figure}[!hbtp]
\begin{center}
\includegraphics[scale=0.3, angle=0]{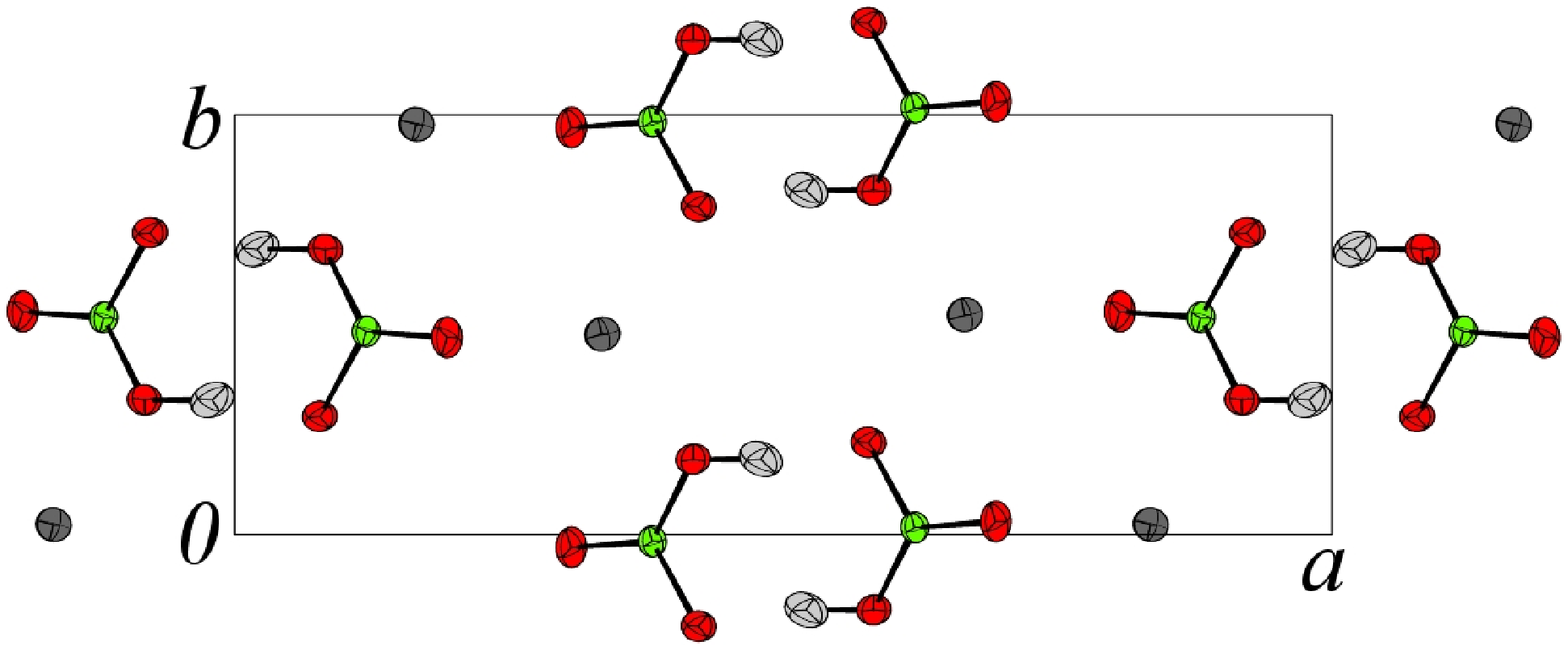}
\makebox[\textwidth]{\ }\\
\includegraphics[scale=0.3, angle=0]{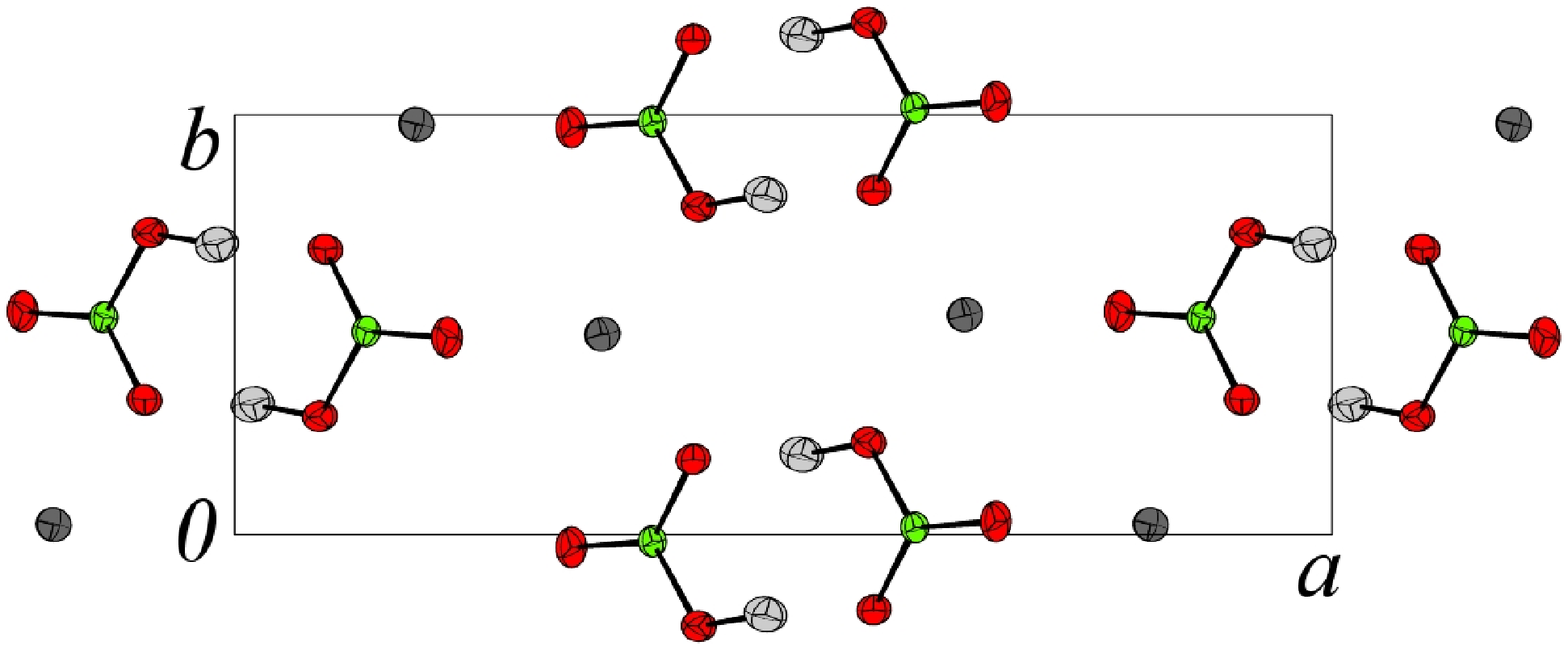}
\end{center}
\caption{\label{fig:2} Projections on the ($a,b$) plane of the $L$ (bottom) and $R$ (top) configurations. }
\end{figure}

From 14 K to 300 K, there is no structural phase transition for either crystal. The increase of the unit cell dimensions and of $R_{\mathrm{OHO}}$ or $R_{\mathrm{ODO}}$ are marginal, while the populations of proton (deuteron) sites change significantly. Below $\approx 150$ K, only those sites corresponding to a unique configuration for dimers (say $L$, see arrows in Fig. \ref{fig:1}) are occupied. At elevated temperatures, less favored $R$ sites, at $\approx 0.6$ \AA\ from the $L$ ones, are progressively populated. The space group symmetry and quantum correlations are unaffected. This accords with a superposition of macroscopically distinct $|L\rangle$ and $|R\rangle$ states corresponding to the $P2_1/a$ configurations sketched in Fig. \ref{fig:2} \cite{FCG2,Fil7,FCG4}. 

\subsection{\label{sec:22}KH$_p$D$_{1-p}$CO$_3$}

An approximately cubic specimen ($3\times 3\times 3$ mm$^3$) was cut from a crystal obtained by cooling slowly a saturated solution in a mixture of H$_2$O ($\approx 75\%$) and D$_2$O ($\approx 25\%$). We utilized the four-circle diffractometer 5C2 based at the Orph\'{e}e reactor (Laboratoire L\'{e}on-Brillouin) \cite{LLB}. The crystal was wrapped in aluminum and maintained at $(300 \pm 1)$ K. The data reduction was carried out with CRYSTALS \cite{WPCBC}. 

Inspection of intensities for absent reflections ($0 k 0$ for $k = 2n+1$ and $h 0 l$ for $h = 2n+1$) confirms the $P2_1/a$ space group assignment. There is no evidence of any symmetry breaking due to a static distribution of protons and deuterons. This information is unique, because neutrons are scattered by nuclei. The same space group assignment, should it be determined with X-ray diffraction, would not rule out a static distribution of isotopes carrying the same number of electrons. 

We refined a structural model comprising four independent sublattices, H$_L$, H$_R$, D$_L$, D$_R$, as: 
\begin{equation}\label{eq:1}
\begin{array}{l}
\mathrm{HD}(p_H, p_D,\rho_{LH}, \rho_{RH}, \rho_{LD}, \rho_{RD}) = \\
p_H [(\rho_{LH}\mathrm{H}_L + \rho_{RH} \mathrm{H}_R] + p_D [\rho_{LD}\mathrm{D}_L + \rho_{RD} \mathrm{D}_R]. 
\end{array}\end{equation}
The sublattices are separate, although they may correspond to identical sites. The coherent scattering lengths were set to either $b_H \approx -3.741$ fm for H$_L$ and H$_R$, or $b_D \approx 6.671$ fm for D$_L$ and D$_R$. Positional and anisotropic thermal parameters for each sublattice, isotope concentrations ($p_H, p_D$) and occupancies ($\rho_{LH},$ $\rho_{RH},$ $\rho_{LD},$ $\rho_{RD}$) were allowed to vary independently from each other. 

In Tables \ref{tab:1} to \ref{tab:3}, the refined parameters are compared to those of the fully hydrogenated analogue at the same temperature. Atoms are labelled H(1), D(1) for $L$ and H(2), D(2) for $R$. The unit cell parameters and the parameters for heavy atom sites are virtually identical. Small differences for proton positions are insignificant compared to the thermal ellipsoids. 

In the isotope mixture, positional and thermal parameters of protons and deuterons at the same sites are identical, so Fig. \ref{fig:1} is representative of the structure with a superposition of indiscernible H and D sublattices. In Table \ref{tab:2}, the estimated concentrations, $p_H = 0.76(2)$ and $p_D = 0.24(2)$ accord with those of the initial water solution and the sum $\rho_{LH} + \rho_{RH} + \rho_{LD} + \rho_{RD} = 0.99(4)$ is self-consistent. 

The occupancy $\rho_{RH} = 0.19(2)$ could be slightly smaller than 0.22(1) for KHCO$_3$, while $\rho_{RD} = 0.15(2)$ could be a little bit greater than 0.12(1) for KDCO$_3$ \cite{TTO2}. A small variation of the effective energy difference between $L$ and $R$ configurations with the isotope content cannot be excluded, but statistical errors are not conclusive. Finally, the O$\cdots$O bond length of 2.587(1) \AA\ is equal to $p_H R_\mathrm{OHO} +p_DR_\mathrm{ODO}$. The lattice of heavy atoms could be also made up of separate sublattices, but we did not pursue this option. 

For the sake of completeness, we tried an alternative model supposing that H and D nuclei merge into hybrid entities [H$_p$D$_{1-p}$] with the averaged scattering length $\bar{b}_p = p b_H + (1-p)b_D$. These hybrids were incorporated in sublattices [H$_p$D$_{1-p}$]$_L$ and [H$_p$D$_{1-p}$]$_R$ as:
\begin{equation}\label{eq:2}
[\mathrm{H}_p\mathrm{D}_{1-p}] (\rho_{L},\rho_{R}) = \rho_{L}[\mathrm{H}_p\mathrm{D}_{1-p}]_{L} + \rho_{R}[\mathrm{H}_p\mathrm{D}_{1-p}]_{R}. 
\end{equation}
In this model the isotope concentration $p$ is not a free parameter. We set $p = 0.76$ and $\bar{b}_{0.76} \approx -1.242$ fm. When $\rho_{L}$ and $\rho_{R}$, as well as positional and thermal parameters for the $L$ and $R$ sublattices, were allowed to vary independently, the refinement did not converge to any stable structure. We imposed further constraints by setting equal thermal parameters for $L$ and $R$. Then, refinement of the remaining 56 free parameters did converge to a stable structure (R = 0.036), roughly similar to the previous one. Therefore, this model is not formally excluded by Bragg diffraction. In the next section, we present theoretical arguments liable to discriminate models (\ref{eq:1}) and (\ref{eq:2}). 

\section{\label{sec:3}Theory}

The theoretical framework is a follow up of that presented in Refs \cite{FCG2,FCG4}. We begin with macroscopic states in pure crystals and then we consider isotope mixtures. 

\subsection{\label{sec:31}The adiabatic separation}

Within the framework of the Born-Oppenheimer approximation, the nuclear Hamiltonian can be partitioned as
\begin{equation}\label{eq:3}
\mathcal{H}_\mathrm{v} = \mathcal{H}_{\mathrm{H}} + \mathcal{H}_{\mathrm{D}} + \mathcal{C}_{\mathrm{H,D}} +\mathcal{H}_{\mathrm{at}}+ \mathcal{C}_{\mathrm{H,D,at}},
\end{equation}
where $\mathcal{H}_{\mathrm{H}}$, $\mathcal{H}_{\mathrm{D}}$, and $\mathcal{H}_{\mathrm{at}}$ represent the sublattices of H$^+$, D$^+$ and heavy atoms, respectively. $\mathcal{C}_{\mathrm{H,D}}$ is the coupling term between $\mathcal{H}_{\mathrm{H}}$ and $\mathcal{H}_{\mathrm{D}}$, while $\mathcal{C}_{\mathrm{H,D,at}}$ couples the three subsystems. 

Vibrational spectra show that H and D states are separable, so $\mathcal{C}_{\mathrm{H,D}}$ can be ignored. The ``hybrid'' model (\ref{eq:2}) leading to an effective oscillator mass $m_\mathrm{H}p + m_\mathrm{D}(1-p)$ and a continuous frequency shift with $p$ is therefore inappropriate. 

For OHO or ODO bonds, coupling terms between O$\cdots$O and H or D degrees of freedom are rather large \cite{SZS,Novak}, hence beyond the framework of the perturbation theory. Yet, quantum dynamics can be rationalized within the framework of the adiabatic separation of $\mathcal{H}_{\mathrm{H}}$ and $\mathcal{H}_{\mathrm{D}}$, on the one hand, and $\mathcal{H}_{\mathrm{at}}$, on the other \cite{FCG2,FCG4}. The observation of separable H and D states is an evidence of the adiabatic separation from heavy atoms. Then, bare protons are fermions while bare deuterons are bosons. 

\subsection{\label{sec:32}Macroscopic states}

Consider a crystal made of $N_\mathrm{a}$, $N_\mathrm{b}$, $N_\mathrm{c}$ ($\mathcal{N}=N_\mathrm{a}N_\mathrm{b}N_\mathrm{c}$) unit cells labelled $j,k,l,$ along crystal axes $(a),$ $(b),$ $(c)$, respectively. The two dimers per unit cell are indexed as $j,k,l$ and $j',k,l$, respectively, with $j = j'$. For centrosymmetric dimers, there is no permanent dipolar interaction, so interdimer coupling terms are negligible \cite{FTP,IKSYBF}. The eigen states of the sublattices of protons or deuterons can be therefore represented with eigen states for isolated dimers 

A dimer (H1,H2) or (D1,D2) is modelled with coupled centrosymmetric collinear oscillators in three dimensions, along coordinates $\{\alpha_{1\mathrm{jkl}}\}$ and $\{\alpha_{2\mathrm{jkl}}\}$ ($\alpha =x,y,z$). The center of symmetry is at $\{\alpha_{0\mathrm{jkl}}\}$. Within the framework of the harmonic approximation, the mass-conserving symmetry coordinates independent of $j, k,l,$ and their conjugated momenta, 
\begin{equation}\label{eq:4}
  \begin{array}{lc}
\alpha_{\mathrm{s}} = \displaystyle{\frac {1} {\sqrt{2}}\left(\alpha_{1} - \alpha_{2} + 2\alpha_0 \right)}, & P_{\mathrm{s}\alpha} = \displaystyle{\frac {1} {\sqrt{2}} \left( P_{1\alpha} - P_{2\alpha} \right)}, \\
    \alpha_{\mathrm{a}} = \displaystyle{\frac {1} {\sqrt{2}}\left(\alpha_{1} + \alpha_{2} \right)}, & P_{\mathrm{a}\alpha} = \displaystyle{\frac {1} {\sqrt{2}}\left( P_{1\alpha} + P_{2\alpha } \right)},
  \end{array}
\end{equation}
lead to uncoupled oscillators at frequencies $\hbar\omega_{\mathrm{s}\alpha}$ and $\hbar\omega_{\mathrm{a}\alpha}$, respectively. The difference $(\hbar\omega_{\mathrm{s}\alpha} - \hbar\omega_{\mathrm{a}\alpha})$ depends on the coupling term (say $\lambda_\alpha$). The wave functions, $\{\Psi_{\mathrm{njkl}}^\mathrm{a}( \alpha_{\mathrm{a}})\}$, $\{\Psi_{\mathrm{n'jkl}}^\mathrm{s}(\alpha_{\mathrm{s}} -\sqrt{2}\alpha_{0})\}$, cannot be factored into wave functions for individual particles, so there is no local information available for these entangled oscillators. Then, the wave functions for protons or deuterons are subjected to the symmetrization postulate of quantum mechanics in different ways \cite{CTDL}. As the wave functions $\{\Psi_{\mathrm{njkl}}^\mathrm{a}( \alpha_{\mathrm{a}})\}$ and $\{\Psi_{\mathrm{n'jkl}}^\mathrm{s}(\alpha_{\mathrm{s}} -\sqrt{2}\alpha_{0})\}$ are unaffected by permutation changing the signs of $\{\alpha_{0\mathrm{jkl}}\}$, they are appropriate for bosons, but not for fermions.

\subsubsection{\label{sec:321}Protons}

In order to antisymmetrize the degenerate ground state with respect to permutation, the wave functions are rewritten as linear combinations 
\begin{equation}\label{eq:5}
\Theta_{0\mathrm{jkl}\pm } = \displaystyle{\frac{1} {\sqrt{2}}} \prod\limits_\alpha \Psi_{0\mathrm{jkl}}^\mathrm{a} (\alpha_{\mathrm{a}}) \left [ \Psi_{0\mathrm{jkl}}^\mathrm{s}(\alpha_{\mathrm{s}} -\sqrt{2}\alpha_{0}) \pm \Psi_{0\mathrm{jkl}}^\mathrm{s} (\alpha_{\mathrm{s}} +\sqrt{2}\alpha_{0}) \right ]
\end{equation}
and the state vectors with singlet-like $|S\rangle$ or triplet-like $|T\rangle$ spin symmetry are: 
\begin{equation}\label{eq:6}
\begin{array}{c}
| \Theta_{0\mathrm{jkl}+} \rangle \otimes |S\rangle ;\\
| \Theta_{0\mathrm{jkl}-} \rangle \otimes |T\rangle .
\end{array} 
\end{equation}
The oscillators are now entangled in position, momentum, and spin. As there is no level splitting, entanglement is energy-free and independent of $\lambda_\alpha$. 

The spatial periodicity of the sublattice leads to collective dynamics and nonlocal observables in three dimensions. With the vibrational wave function for the unit cell $j,\ k,\ l,$ namely $\Xi_{0\mathrm{jkl}\tau } = \Theta_{0\mathrm{jkl}\tau } +\tau \Theta_{0\mathrm{j'kl}\tau }$, where $\tau =$ ``$+$'' or ``$-$'' for singlet-like or triplet-like symmetry, respectively, we write phonon waves as 
\begin{equation}\label{eq:7}
\Xi_{0\tau} (\mathbf{k})= \displaystyle{\frac{1}{\sqrt {\mathcal{N}}}}  \sum\limits_{\mathrm{l} = 1}^{\mathrm{N_c}} \sum\limits_{\mathrm{k} = 1}^{\mathrm{N_b}} \sum\limits_{\mathrm{j} = 1}^{\mathrm{N_a}} \Xi_{0\mathrm{jkl}\tau} \exp(i\mathbf{k\cdot L}), 
\end{equation}
where $\mathbf{k}$ is the wave vector, $\mathbf{L}  = j \mathbf{a} + k \mathbf{b} + l \mathbf{c}$ is the lattice vector, $\mathbf{a}$, $\mathbf{b}$, $\mathbf{c}$, are the unit cell vectors. Then, antisymmetrization leads to 
\begin{equation}\label{eq:8}
\mathbf{k\cdot L} \equiv 0 \mathrm{\ modulo\ } 2\pi. 
\end{equation}
This means that there is no phonon (no elastic distortion) in the ground state (``super-rigidity'') \cite{FCG2}. The lattice state vectors 
\begin{equation}\label{eq:9}
\begin{array}{c}
|\mathrm{H}_+\rangle = \left | \Xi_{0 +} (\mathbf{k = 0}) \right \rangle \otimes |S \rangle \\
|\mathrm{H}_-\rangle = \left | \Xi_{0 -} (\mathbf{k = 0}) \right \rangle \otimes |T \rangle 
\end{array}
\end{equation}
are superposed in the ground state as:
\begin{equation}\label{eq:10}
\begin{array}{c}
\sqrt{\mathcal{N}} | \Xi_{0 +} (\mathbf{k = 0}) \rangle \otimes |S \rangle;\\ 
\sqrt{\mathcal{N}} | \Xi_{0 -} (\mathbf{k = 0}) \rangle \otimes |T \rangle. 
\end{array}
\end{equation}
These states are intrinsically steady against decoherence. The main source of disentanglement is the thermal bath but, even at room temperature, the population of the first excited state ($< 1\%$ for $\gamma$OH $\approx 1000$ \cm) is negligible. 

\subsubsection{\label{sec:322}Deuterons}

For bosons, the wave functions for a dimer 
\begin{equation}\label{eq:11}
\Theta_{0\mathrm{jkl}} = \prod\limits_\alpha \Psi_{0\mathrm{jkl}}^\mathrm{a} (\alpha_{\mathrm{a}}) \Psi_{0\mathrm{jkl}}^\mathrm{s}(\alpha_{\mathrm{s}} -\sqrt{2}\alpha_{0}) ,
\end{equation}
are symmetric with respect to particle permutation. As there is no significant spin-spin coupling compared to the thermal bath, the spin of a pair is statistically zero, so there is no correlation at all. Phonon states can be written as 
\begin{equation}\label{eq:12}
|\mathrm{D}_{\mathbf{k}\pm}\rangle= \displaystyle{\frac{1}{\sqrt {\mathcal{N}}}}  \sum\limits_{\mathrm{l} = 1}^{\mathrm{N_c}} \sum\limits_{\mathrm{k} = 1}^{\mathrm{N_b}} \sum\limits_{\mathrm{j} = 1}^{\mathrm{N_a}} |\Xi_{0\mathrm{jkl}\pm}\rangle \exp(i\mathbf{k\cdot L}), 
\end{equation}
where $\Xi_{0\mathrm{jkl}\pm} = \Theta_{0\mathrm{jkl}} \pm \Theta_{0\mathrm{j'kl}}$. In contrast to the fermion case, there is no super-rigidity. Needless to say, the H and D sublattices have the same number of degrees of freedom (12 per unit cell), but the symmetrization postulate shrinks the size of the allowed Hilbert space from $\sim 12^\mathcal{N}$ for bosons to $\sim 12\mathcal{N}$ for fermions. 

\subsubsection{\label{sec:323}Isotope mixtures}

The sublattice at room temperature can be represented with a superposition of separable proton and deuteron states, each of them being a superposition of $L$ and $R$ states, as: 

\begin{equation}\label{eq:13}
p_H^{1/2} [(\rho_{LH}^{1/2}|\mathrm{H}_{\pm}\rangle_L + \rho_{RH}^{1/2} |\mathrm{H}_{\pm}\rangle_R] + p_D^{1/2} [\rho_{LD}^{1/2}|\mathrm{D}_{\mathbf{k}L\pm}\rangle_L + \rho_{RD}^{1/2} |\mathrm{D}_{\mathbf{k}R\pm}\rangle_R]; 
\end{equation}

\subsection{\label{sec:33}Neutron diffraction}

Neutrons (spin $1/2$) are unique to probing the spin-symmetry of proton states (\ref{eq:9}). However, the spin entanglement is extremely fragile, so only ``noninvasive'' experiments, free of measurement-induced decoherence, are appropriate \cite{LG}. This means: (i) no energy transfer; (ii) no spin-flip; (iii) particular values of the neutron momentum transfer vector $\Q$ \cite{Q} preserving the super-rigidity. This is realized when the components $Q_\mathrm{x}$, $ Q_\mathrm{y}$, $Q_\mathrm{z},$ match a node of the reciprocal sublattice of protons \cite{FCG2,FCG4}. 

With a reactor-based diffractometer, these conditions are not realized at the measured Bragg peaks. The spin-symmetry is destroyed, the super-rigidity is lost, and the final proton states $|\mathrm{H}_{\mathbf{k}\pm}\rangle$ are analogous to boson states (\ref{eq:12}). The structural model (\ref{eq:1}) is therefore consistent with diffraction by an initial superposition state (\ref{eq:13}) and one of the final states $|\mathrm{H}_{\mathbf{k}\pm}\rangle_L,$ $|\mathrm{H}_{\mathbf{k}\pm}\rangle_R,$ $|\mathrm{D}_{\mathbf{k}\pm}\rangle_L,$ $|\mathrm{D}_{\mathbf{k}\pm}\rangle_R$. The final state of a particular event is unknown because the four states are degenerate in reciprocal space. The probabilities of these scattering events are given by (\ref{eq:1}), in accordance with (\ref{eq:13}). 

\section{Conclusion}

The space group symmetry of KH$_{0.76}$D$_{0.24}$CO$_3$ determined by single-crystal neutron diffraction shows that H and D nuclei are not individual and mutually exclusive particles located at definite sites. There is no disorder, either static or dynamic. Bragg peaks are best fitted with separate H and D sublattices of identical sites. Each sublattice is further made up of distinct $L$ and $R$ sublattices. This structure accords with a superposition of separable macroscopic states. Further measurements should provide information as to whether the spin-symmetry of proton states survives in isotope mixtures. 

There is now a substantial amount of experimental data showing that crystals of KHCO$_3$ and isotope derivatives are macroscopic quantum objects on the scale of Avogadro's constant and room temperature. The wave, or matter-field, representation is imposed by the electronic structure insensitive to isotope substitution. The theoretical framework is based on fundamental laws of quantum mechanics. It is free of any ad hoc hypothesis or parameter. There is every reason to suppose that macroscopic states should occur in many systems. Neutron diffraction by partially deuterated hydrogen bonded crystals is well-suited to evidencing the nonlocal character of protons and deuterons, as opposed to disorder.

\clearpage

\begin{table}[!hbtp]
\caption{\label{tab:1} Neutron single crystal diffraction data and structure
refinement at 300 K. $\lambda$ = 0.8305 \AA, space group $P 2_1/a$. The variance for the last digit is given in parentheses.}
\begin{tabular}{lll}
\hline
 Crystal data  & KH$_{0.76}$D$_{0.24}$CO$_3$ & KHCO$_3$ \cite{FCG2} \\
\hline
$a$(\AA) & 15.180(1) & 15.180(1) \\
$b$(\AA) & 5.620(1)  & 5.620(4) \\
$c$(\AA) & 3.710(1)  & 3.710(4) \\
$\beta$  & 104.67(1)$^\circ$ & 104.67(5)$^\circ$ \\
Volume   & 306.2  & 306.2\\
Reflections measured & 1288  & 1731 \\
Independent reflections & 1011 & 1475 \\
Reflections used      & 811 & 1068 \\
$\sigma$(I) limit & 3.00 & 3.00\\
Refinement on F & & \\
R-factor & 0.026 & 0.035 \\
Weighted R-factor & 0.038 & 0.034 \\
Number of parameters & 77 & 56 \\
Goodness of fit & 1.043 & 1.025 \\
Extinction & 121.8(40) & 3260(100)\\
\hline
\end{tabular}
\end{table}

\begin{table}
\caption{\label{tab:2} Atomic positions, isotropic temperature factors and
site occupancies for KH$_{0.76}$D$_{0.24}$CO$_3$ (first lines) 
and KHCO$_3$ (second lines) \cite{FCG2} at 300 K. The variance for the last digit is given in
parentheses.}
\begin{tabular}{llllll}
\hline
 Atom & $x/a$ & $y/b$ & $z/c$ & U(iso)(\AA$^2$) & Occupancy\\
\hline
K(1) & 0.16534(6) & 0.0242(2)& 0.2956(2) & 0.0232 & 1.000 \\
     & 0.16534(6) & 0.0230(2)& 0.2952(2)   & 0.0213 & 1.000 \\
C(1) & 0.11963(3) & 0.51735(8) & -0.1443(1)& 0.0185  & 1.000 \\
     & 0.11957(3)  & 0.51636(8) & -0.1444(1) & 0.0164 & 1.000 \\
O(1) & 0.19345(4)  & 0.5321(1)  & 0.0947(2) & 0.0279 & 1.000 \\
     & 0.19344(4)  & 0.5307(1) & 0.0946(2) & 0.0259 & 1.000 \\
O(2) & 0.08237(4)  & 0.3220(1) & -0.2734(2) & 0.0260 & 1.000 \\
     & 0.08237(4)  & 0.3206(1)  & -0.2735(2) & 0.0238 & 1.000 \\
O(3) & 0.07737(4)  & 0.7195(1) & -0.2741(2) & 0.0266 & 1.000 \\
     & 0.07745(4)  & 0.7186(1)  & -0.2743(2)& 0.0243 & 1.000 \\
H(1) & 0.0158(3)  & 0.6897(3) & -0.4492(3)  & 0.0310 & 0.643(7) \\
     & 0.01631(12) & 0.6905(2)   & -0.4491(5)  & 0.0330 & 0.823(4) \\
D(1) & 0.0158(3)  & 0.6897(3) & -0.4492(3)  & 0.0310 & 0.200(6) \\
     & --- & --- & --- & --- & --- \\
H(2) & -0.019(2)   & 0.686(2)    & -0.560(2)    & 0.0324  & 0.12(1) \\
     & -0.0207(6) & 0.680(1)   & -0.563(2)   & 0.0338  & 0.177(4) \\
D(2) & -0.019(2)   & 0.686(2)    & -0.560(2)    & 0.0324  & 0.03(1) \\
     & --- & --- & --- & --- & --- \\
\hline
\end{tabular}
\end{table}

\begin{table}
\caption{\label{tab:3} Thermal parameters in \AA$^2$ units for KH$_{0.76}$D$_{0.24}$CO$_3$ (first lines) and KHCO$_3$ (second lines) \cite{FCG2} at 300 K. The variance for the last digit is given in parentheses. These parameters account for the contribution of each atom to Bragg's peak intensities through the thermal factor $T^{at}  = \exp\left [ -2\pi^2 (U_{11}^{at}\textrm{h}^2a^{*2} + U_{22}^{at}\textrm{k}^2b^{*2} + U_{33}^{at}\textrm{l}^2c^{*2} + 2U_{12}^{at}\textrm{h}a^*\textrm{k}b^* + 2U_{23}^{at}\textrm{k}b^*\textrm{l}c^* + U_{31}^{at}\textrm{l}c^*\textrm{h}a^* ) \right ],$ where $a^*$, $b^*$, $c^*$, are the reciprocal lattice parameters and h, k, l, the indexes in reciprocal space.}
\begin{tabular}{lllllll}
\hline
 Atom & U$_{11}$ & U$_{22}$ & U$_{33}$ & U$_{23}$ & U$_{13}$ & U$_{12}$ \\
\hline
K(1)  & 0.0265(4)   & 0.0231(4)   & 0.0181(3)    & 0.0001(3)  & 0.0018(3)  & 0.0006(3) \\
      & 0.0240(3)   & 0.0222(3)   & 0.0159(3)    & -0.0004(3)   & 0.0017(2) & 0.0009(3) \\
C(1)  & 0.0178(2) & 0.0194(2) & 0.0171(2)  & -0.0003(1)   & 0.00207(13)   & -0.0003(1) \\
      & 0.0150(2) & 0.0182(2) & 0.0147(2)  & -0.0000(1) & 0.0016(2)  & -0.0003(1) \\
O(1)  & 0.0213(3) & 0.0317(3) & 0.0256(3)  & -0.0016(2) & -0.0034(2) & -0.0000(2) \\
      & 0.0177(2)   & 0.0316(3)   & 0.0232(2)    & -0.0015(2)   & -0.0042(2) & 0.0001(2) \\
O(2)  & 0.0260(3) & 0.0189(2) & 0.0277(3)  & -0.0006(2) & -0.0020(1)  & -0.0031(2) \\
      & 0.0226(2)   & 0.0179(2)   & 0.0255(3)    & -0.0009(2) & -0.0037(2)  & 0.0010(2) \\
O(3)  & 0.0251(3) & 0.0190(2) & 0.0303(3)  & 0.0003(2) & -0.0028(2)  & -0.0005(2) \\
      & 0.0231(2) & 0.0176(2) & 0.0272(3)    & 0.0007(2)  & -0.0030(2)    & -0.0006(2) \\
H(1)/D(1)  & 0.030(6)   & 0.027(6)   & 0.032(7)    &-0.000(6)   & -0.001(6)     & -0.000(6) \\
      & 0.0349(2)  & 0.0236(2) & 0.0369(2)  & -0.0024(2)  & 0.0023(2)    & -0.0021(2) \\
H(2)/D(2)  & 0.045(13) & 0.015(8) & 0.036(13) & 0.007(8) & 0.01(1) & 0.005(8) \\
      & 0.0347(2)   & 0.0237(2)   & 0.0367(2)    & -0.0068(2)   & -0.0023(2)    & -0.0059(2) \\
\hline
\end{tabular}
\end{table}

\clearpage

\begin{thebibliography}{10}
\bibitem{Q}
$\mathbf{Q} = \mathbf{k}_i - \mathbf{k}_f$, where $\mathbf{k}_i$ and
  $\mathbf{k}_f$ are the initial and final wave vectors, respectively.
\end{thebibliography}
\end{document}